\documentclass[journal]{IEEEtran}
\pdfoutput=1
\usepackage{amsmath,amsfonts}
\usepackage{algorithm}
\usepackage{array}
\usepackage{textcomp}
\usepackage{url}
\usepackage{verbatim}
\def\BibTeX{{   B\kern-.05em{\sc i\kern-.025em b}\kern-.08em
    T\kern-.1667em\lower.7ex\hbox{E}\kern-.125emX}}
\usepackage{balance}
\usepackage{amssymb}

\usepackage{setspace} 
\usepackage{stfloats}
\usepackage{graphicx}
\usepackage{bm}
\usepackage{cite}

\usepackage[normalem]{ulem}
\usepackage{color}
\usepackage{cases}
\usepackage{mathtools}
\usepackage{mathrsfs}
\usepackage{indentfirst}
\usepackage{utfsym}
\usepackage{algorithmic}
\usepackage{enumerate}
\usepackage[doipre={doi:~}]{uri}

\usepackage{booktabs}
\usepackage{multicol}
\usepackage{multirow}
\usepackage{diagbox}
\usepackage{txfonts}    % Times Roman
\usepackage{tikz}

\usepackage[colorlinks=true,linkcolor=black,citecolor=black,urlcolor=black]{hyperref}
\usepackage[caption=false,font=footnotesize,labelfont=sf,textfont=rm]{subfig}
\usepackage{makecell}

\usepackage{threeparttable}
\usepackage{orcidlink}

\begin{document}
\bibliographystyle{IEEEtran} 
\title{Dual-Qubit Hierarchical Fuzzy Neural Network for Image Classification: Enabling Relational Learning via Quantum Entanglement}
\author{Wenwei Zhang$^{\orcidlink{0009-0003-5734-7670}}$, Jintao Wang$^{\orcidlink{0009-0003-4712-5353}}$, Tianyu Ye$^{\orcidlink{0000-0002-5581-9895}}$ and Changgeng Liao$^{\orcidlink{0000-0003-0104-3814}}$ 
\thanks{
	This work was supported in part by the National Natural Science Foundation of China (Grants No. 12004336, No. 12075205,and No. 62071430); in part by the International Exchanges 2022 Cost Share (NSFC) from The Royal Society through Grant No. IEC\textbackslash NSFC\textbackslash 223131; in part by the Fundamental Research Funds for the Provincial University of Zhejiang (Grant No. XRK23006); and in part by Funds from the China Scholarship Council. (Corresponding author: Chang-Geng Liao; Tianyu Ye.)
	
	Wenwei Zhang, Jintao Wang and Tianyu Ye are with the School of Information and Electronic Engineering (Sussex Artificial Intelligence Institute), Zhejiang Gongshang University, Hangzhou, Zhejiang 310018, China (e-mail: 24020090040@pop.zjgsu.edu.cn; 1811080102@pop.zjgsu.edu.cn; yetianyu@zjgsu.edu.cn).
	
	Changgeng Liao is with the School of Information and Electronic Engineering (Sussex Artificial Intelligence Institute), Zhejiang Gongshang University, Hangzhou, Zhejiang 310018, China and also with Department of Physics and Astronomy, University of Sussex, Brighton BN1 9QH, United Kingdom (E-mail: cgliao@zjgsu.edu.cn).
	
	Our code is available at \href{https://github.com/Comets9224/DQ-HFNN.git}{https://github.com/Comets9224/DQ-HFNN.git}.
}
}

%\markboth{Journal of \LaTeX\ Class Files,~Vol.~X, No.~X, July~2025}{}

\maketitle
\begin{abstract}
	Classical deep neural network models struggle to represent data uncertainty and capture dependencies between features simultaneously, especially under fuzzy or noisy conditions. Although a quantum-assisted hierarchical fuzzy neural network (QA-HFNN) was proposed to learn fuzzy membership for each feature, it cannot model dependencies between features due to its single-qubit encoding. To address this, this paper proposes a dual-qubit hierarchical fuzzy neural network (DQ-HFNN), encoding feature pairs onto a pair of entangled qubits, which extends the single-feature fuzzy model to a joint fuzzy representation. By introducing quantum entanglement, the dual-qubit circuit can encode non-classical correlations, enabling the model to directly learn relationship patterns between feature pairs. Experiments on benchmarks show that DQ-HFNN demonstrates higher classification accuracy than QA-HFNN, as well as classical deep learning baselines. Furthermore, ablation studies after controlling for circuit depth and parameter counts show that the performance gain mainly stems from the relational modeling capability enabled by entanglement rather than enhanced expressivity. The proposed DQ-HFNN model exhibits high parameter efficiency and fast inference speed. Experiments under noisy conditions suggest that it is robust against noise and has the potential to be implemented on noisy intermediate-scale quantum devices.
\end{abstract}
\section{Introduction}
Classification problems usually involve various forms of uncertainty when it comes to the real world, such as linguistic ambiguity, gradual transitions, or imprecise measurements. The concept of fuzzy logic proposed by Zadeh~\cite{zadeh1965fuzzy} aims to handle these uncertain problems. Compared with classical binary logic (e.g., the weather is either hot or not hot), fuzzy logic allows an element to belong to a set with partial membership (e.g., 25°C can be considered 0.7 warm and 0.3 hot), referred to as membership. By assigning a value with a domain ranging from 0 to 1, the membership function quantifies the degree to which an element belongs to a set. fuzzy neural networks (FNNs) are the combination of fuzzy logic and classical neural networks, which have been proven effective for those complex data-driven tasks. To be specific, a hierarchical fused fuzzy network is developed for data classification~\cite{deng2016hierarchical}; a fuzzy attention mechanism is applied to improve pneumonia detection in medical imaging~\cite{roy2024fa}; an optimized fuzzy deep learning model is created with genetic algorithms~\cite{yazdinejad2023optimized}; and a fuzzy-guided multi-granular network is designed for histopathological image analysis~\cite{ding2024fmdnn}, among others. However, these FNN models are highly dependent on artificial construction, whose shallow feature representations limit their performance on high-dimensional data.

By constructing a multi-layer network structure, deep learning can automatically learn hierarchical representations from raw data~\cite{lecun2002gradient}, making it the mainstream paradigm of machine learning. For example, convolutional neural network (CNN) architectures~\cite{he2016deep} can effectively extract spatial local patterns and combine them into high-level features. The stacked convolutional and pooling layers of CNN can detect local patterns (such as edges and textures) and gradually build more abstract representations. Although deep learning models can effectively extract features, they tend to focus on the features themselves rather than the relationships between features.

Since the meaning of features is typically defined by their context, analyzing individual features is not sufficient enough to capture complete information. For instance, the semantic meaning of a single pixel cannot be determined, as its explanation depends on the spatial and contextual relationship with adjacent pixels. The goal of relational learning is to understand the dependencies among features. The Transformer architecture~\cite{vaswani2017attention} proposed in 2017 employs a self-attention mechanism to capture these relationships. Before this, graph neural networks (GNNs)~\cite{battaglia2018relational} are explicitly designed to learn relationships from predefined graph structures. However, GNNs require the prior knowledge of the relationship graph, or must infer it jointly with the task, which introduces additional complexity~\cite{jiang2019semi}. The representational ability of classical methods is limited by memory resources and computing speed. Explicitly representing the complex distribution of joint features costs high computing resources, which leads to a bottleneck in representation capabilities. This limitation motivates the exploration of computational models operating in higher-dimensional representation spaces.

Quantum computing can achieve parallel computing due to its entanglement and superposition characteristics, which has demonstrated performance surpassing that of classical computers in specific problems~\cite{shor1994algorithms,grover1996fast}. Different from classical bits, qubits exist in the superposition state of 0 and 1 states with a certain probability, naturally providing a continuous space for the fuzzy membership function representing features. Recently, Wu et al. proposed a quantum-assisted hierarchical fuzzy neural network (QA-HFNN) for image classification, which utilizes quantum neural networks (QNNs) composed of single-qubit circuits to capture fuzzy features of data~\cite{wu2024quantum, tiwari2024quantum}. However, this single-qubit encoding method can only handle each $f(x_i)$ feature individually; it does not consider and is unable to process the context information from the dependencies between features.

In this paper, we propose a dual-qubit hierarchical fuzzy neural network (DQ-HFNN), encoding feature pairs $(x_i, x_j)$ onto a pair of entangled qubits. The proposed DQ-HFNN extends the single-qubit system to a dual-qubit system, and introduces quantum entanglement into the trainable circuit. Entanglement~\cite{babu2025entanglement} is a resource that can model relationship by creating non-classical correlations, and enhance feature representations. Thus the model can learn the joint membership function $f(x_i, x_j)$ between two features, which represents relationships between feature pairs rather than individual features.

The main contributions of this paper are as follows:

\begin{enumerate}
	\item We propose a novel DQ-HFNN model by using entangled qubits to model relationships between feature pairs. Compared with the QA-HFNN based on single-qubits, the proposed DQ-HFNN with entangled qubits achieves higher accuracy across multiple benchmarks.
	\item We design systematic ablation studies, and the results show that the performance gains are mainly attributed to the relational modeling capability enabled by entanglement, rather than from increased expressivity.
	\item Finally, we conduct a comprehensive analysis of the efficiency of DQ-HFNN, demonstrating its advantages in terms of parameter count, inference speed, and trainability. Combined with its robustness under quantum noise, these advantages indicate the feasibility of its deployment on noisy intermediate-scale quantum (NISQ) devices~\cite{preskill2018quantum}.
\end{enumerate}

The remainder of this paper is organized as follows. Section II reviews the related work in fuzzy logic, deep learning, and quantum computing. Section III details the DQ-HFNN architecture and methodology. Section IV presents our experimental setup, results, and analysis. Finally, Section V concludes the paper and outlines future research directions. The source code will be made available upon acceptance.

\section{Related Work}

Our work integrates concepts from three complementary fields: fuzzy logic provides tools for modeling uncertainty; deep learning enables automatic feature and relation extraction; and quantum computing offers richer representational capacity.

\subsection{The Evolution of Fuzzy Logic in Machine Learning}

Zadeh introduced fuzzy logic to handle fuzzy and uncertain problems. Fuzzy logic can be used to model the imprecision inherent in human reasoning systems and complex systems~\cite{zadeh1965fuzzy, zadeh1973outline}. Its core component is the membership function, which allows an element to belong partially to a set. It reflects the degree to which an element belongs to a set, and provides a more nuanced representation than the binary logic in classical set theory. This representation introduced by fuzzy logic has driven the development of FNNs. An early FNN model utilizing fuzzy logic is the adaptive-network-based fuzzy inference system, which demonstrated that fuzzy rules and membership functions can be learned directly from data~\cite{jang1993anfis, lin1991neural}.

This paradigm of combining classical and fuzzy logic has further led to the emergence of hybrid fuzzy-deep architectures. Hybrid architectures can embed fuzzy reasoning within modern neural networks, addressing the specific challenge that data in the medical domain is typically complex and uncertain. Recent applications include using fuzzy attention modules to improve pneumonia detection in chest X-rays~\cite{roy2024fa}, designing fuzzy-guided multi-granular networks for histopathological image analysis~\cite{ding2024fmdnn}, and developing optimized fuzzy deep learning models for monitoring chronic cardiovascular disease~\cite{mohamed2025deep}. However, this also has limitations. Many fuzzy-neural models still depend on shallow architectures and hand-crafted features, limiting their scalability to high-dimensional raw data.
\subsection{From Feature Extraction to Relational Learning in Deep Learning}

Deep learning can learn hierarchical features from complex data such as images, which is beneficial for addressing the limitations of hybrid fuzzy-deep architectures. The CNN is an important network architecture in deep learning. The stacked convolutional and pooling layers of CNN can detect local patterns (such as edges and textures) and effectively extract features automatically~\cite{lecun2002gradient, he2016deep}. However, it tends to focus on the features themselves rather than the relationships between features. As the field matured, it became clear that understanding the context and relationships between features is as important as the features themselves. This realization spurred the development of relational learning.

Two important classical deep learning paradigms have been used for relational learning. The first is the Transformer architecture, which computes pairwise dependencies through self-attention to dynamically learn feature relationships~\cite{vaswani2017attention, dosovitskiy2020image}. The second is GNN, which operates on data explicitly structured as graphs of nodes and edges, propagating information based on these typically predefined relationships~\cite{battaglia2018relational}. The frontier of this research continues to advance rapidly, with recent work focusing on enabling GNNs to operate directly on complex relational databases~\cite{chen2025relgnn} and even integrating them with Large Language Models to enhance their reasoning capabilities over structured data~\cite{wu2025large}. However, despite certain successes, classical architectures face practical representational bottlenecks: explicitly modeling high-order joint distributions becomes expensive as scale grows. These limitations motivate further exploration of computational paradigms that can provide richer representational capacity.
\subsection{Quantum Computing for Enhanced Representation}
Quantum computing provides a possible solution to this representational challenge. By leveraging principles such as superposition and entanglement, quantum systems operate in an exponentially large Hilbert space and can provide a continuous function space, offering in principle more complex representational capacity compared to classical computing. The potential of quantum machine Learning (QML) was recognized in early algorithm proposals, such as the Quantum Support Vector Machine~\cite{rebentrost2014quantum}. In the current era of near-term quantum devices, the dominant approach is the hybrid quantum-classical model, where parameterized quantum circuit (PQC) is trained using classical optimization methods.

\begin{figure*}[!t]
	\centering
	\includegraphics[scale=0.90]{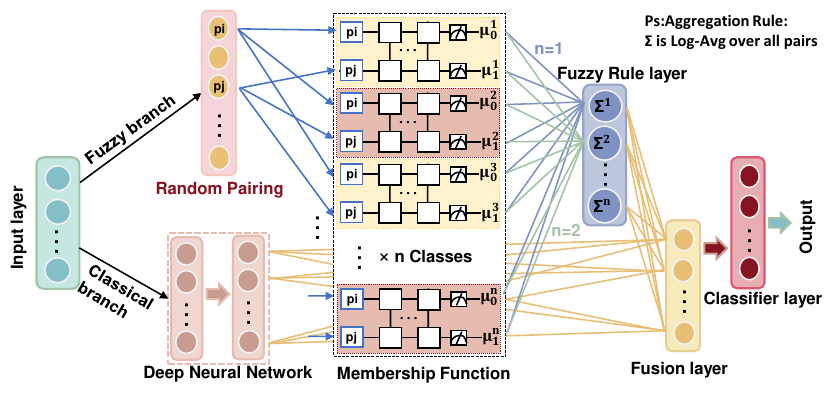}
	\caption{The overall architecture of the proposed dual-qubit hierarchical fuzzy neural Network (DQ-HFNN). The model consists of two parallel branches: a quantum fuzzy branch (top path) and a classical deep neural network (DNN) branch (bottom path). The quantum branch employs a pairing strategy to select feature pairs (e.g., pixels $p_i, p_j$), which are processed by class-specific dual-qubit circuits. The outputs are then aggregated by a fuzzy rule layer ($\Sigma^n$). Concurrently, the classical branch extracts high-level features using a standard DNN. Features from both branches are integrated in a fusion layer and passed to a final classifier for prediction.}
	\label{fig:architecture}
\end{figure*}

This quantum paradigm has been successfully applied to fuzzy logic. Combining the hybrid quantum-classical model with fuzzy logic has created the emerging field of quantum fuzzy neural networks (QFNNs). Existing work has established that a single-qubit, with its ability to exist in a continuous superposition of states, can act as a highly expressive membership function for an individual feature under standard PQC expressibility metrics~\cite{wu2024quantum}. This approach has shown promise in diverse applications from image classification to multimodal sentiment and sarcasm detection~\cite{tiwari2024quantum}, and has even been extended to frameworks such as quantum federated learning for privacy protection~\cite{qu2024quantum}.

However, existing QFNNs primarily rely on single-qubit representations, where each feature is encoded independently and the architecture does not employ entanglement. While this approach is effective for modeling uncertainty, such separable architectures cannot exploit entanglement, a known key resource for representing non-classical correlations between features. Entanglement creates correlations between qubits that have no classical equivalent and has been identified as a key resource for achieving quantum advantage and enabling enhanced feature representations~\cite{babu2025entanglement,huang2022quantum}. These observations point toward a natural progression: extending quantum-fuzzy models from single-qubit to multi-qubit circuits capable of leveraging entanglement.

In summary, while classical deep learning excels at feature and relational learning, and single-qubit QFNNs effectively model feature-wise uncertainty, to the best of our knowledge, no existing framework combines fuzzy reasoning with entanglement-based relational modeling~\cite{yao2025hqfnn,tiwari2024quantum,yerkin2025fuzzy}. The DQ-HFNN proposed in this work is designed to fill this gap, proposing a framework where entanglement is not an incidental byproduct, but the central mechanism for performing relational learning.
\section{PROPOSED METHOD}
\label{sec:proposed_method}

\begin{figure*}[!t]
	\centering
	\includegraphics[scale=0.90]{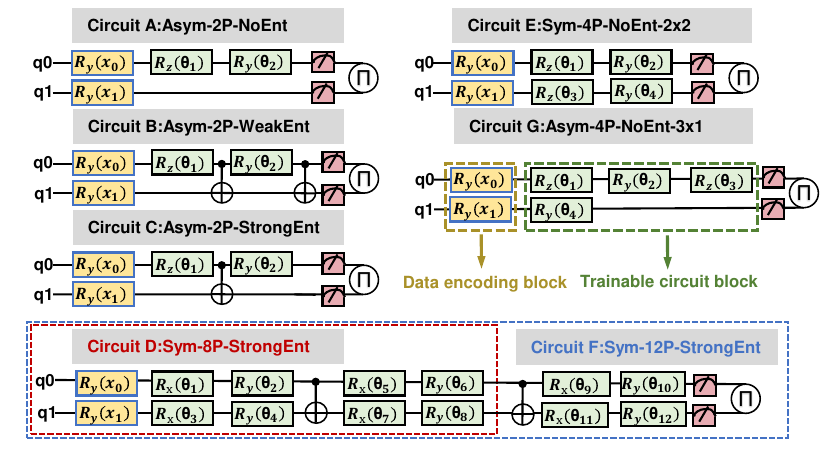}
	\caption{Quantum circuit designs for the trainable block $\boldsymbol{U}_q(\boldsymbol{\theta})$ used in the ablation study. The figure illustrates seven distinct dual-qubit architectures (A--G) with varying degrees of parameterization (P), symmetry (Sym/Asym), and entanglement (Ent). The yellow boxes represent the data encoding block ($\boldsymbol{U}_{\text{enc}}$), where input features $x_0$ and $x_1$ are encoded using $\boldsymbol{R}_y$ gates. The green boxes represent the trainable circuit blocks ($\boldsymbol{U}_q$), which consist of single-qubit rotation gates (e.g., $\boldsymbol{R}_x, \boldsymbol{R}_z$) and entangling CNOT gates. The final block ($\Pi$) denotes the joint measurement of Pauli-Z operators on both qubits.}
	\label{fig:circuit}
\end{figure*}

\begin{figure}[!t]
	\centering
	\includegraphics[scale=0.95]{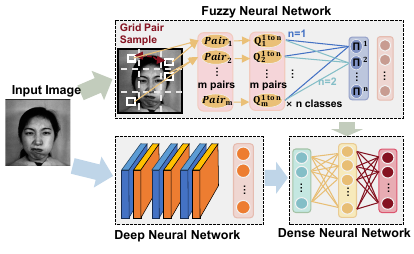}
	\caption{Illustration of the end-to-end workflow of the DQ-HFNN model on a sample image from the JAFFE~\cite{lyons1998coding,lyons2021excavating} dataset. The model operates in two parallel streams. In the quantum stream (top), a grid-based pairing strategy is applied to the input image, sampling pixel pairs (e.g., $\textit{Pair}_1, \dots, \textit{Pair}_m$) to capture multi-scale spatial relationships. These pairs are then processed by the quantum fuzzy logic branch. In the classical stream (bottom), the full image is processed by a CNN to extract semantic features. Finally, the outputs from both streams are fused and classified.}
	\label{fig:jaffe_grid}
\end{figure}

As shown in Figure~\ref{fig:architecture}, the DQ-HFNN comprises two parallel branches: a quantum fuzzy branch that processes feature pairs, and a classical DNN branch that extracts high-level semantic features. The two representations are projected into a shared space and fused before being fed into the final classifier.

\subsection{Quantum Fuzzy Logic Branch}
\label{subsec:quantum_fuzzy_branch}

The quantum branch is the core innovation of our model, designed to capture relational information through a novel fuzzy logic framework that leverages quantum entanglement.
\subsubsection{From Fuzzy Sets to Joint Fuzzy Relations}
\label{subsubsec:fuzzy_to_joint}

Classical fuzzy systems, as first proposed by Zadeh~\cite{zadeh1965fuzzy}, define a fuzzy set $\mathcal{A}$ on a universe of discourse $\mathcal{X}$ as $\mathcal{A} = \{(x_k, \mu_A(x_k)) \mid x_k \in \mathcal{X}\}$, where $\mu_A: \mathcal{X} \to [0, 1]
$ is a membership function for an individual element $x_k$. Our model extends this concept to operate on pairs of features $(x_i, x_j)$. Consequently, the universe of discourse is expanded to the Cartesian product $\mathcal{X} \times \mathcal{X}$.

In classical fuzzy relation theory~\cite{zadeh1971similarity}, a binary fuzzy relation is defined as
\begin{equation}
	\label{eq:classical_fuzzy_relation}
	\mathcal{R} = \{((x_i, x_j), \mu_R(x_i, x_j)) \mid (x_i, x_j) \in \mathcal{X} \times \mathcal{X}\},
\end{equation}
where $\mu_R: \mathcal{X} \times \mathcal{X} \to [0, 1]$ is a scalar-valued membership function. However, to fully exploit the information encoded in our dual-qubit quantum system, we adopt a vector-valued representation inspired by interval-valued fuzzy sets~\cite{zadeh1975concept} and intuitionistic fuzzy sets~\cite{atanassov1999intuitionistic}, which extend classical fuzzy theory to capture additional dimensions of uncertainty through multi-valued membership functions.

Specifically, we define a quantum-enhanced fuzzy relation $\mathcal{R}$ as
\begin{equation}
	\label{eq:fuzzy_relation}
	\mathcal{R} = \{((x_i, x_j), f_{\boldsymbol{\theta}}(x_i, x_j)) \mid (x_i, x_j) \in \mathcal{X} \times \mathcal{X}\},
\end{equation}
where $f_{\boldsymbol{\theta}}(x_i, x_j) = [\mu_0(x_i, x_j), \mu_1(x_i, x_j)] \in [0, 1]^2$ is a vector-valued joint membership function. Each component $\mu_k$ corresponds to the fuzzy degree extracted from one of the two entangled qubits, representing complementary aspects of the feature relationship. These dual measurements are subsequently aggregated using the log-geometric mean operator (\ref{eq:fuzzy_aggregation}), which is the logarithmic form of the geometric mean\cite{zimmermann1980latent}.

\subsubsection{Quantum Implementation of the Joint Membership Function}
\label{subsubsec:quantum_implementation}

The joint membership function $f_{\boldsymbol{\theta}}(x_i, x_j)$ is realized using a PQC. Prior to encoding, feature values are linearly normalized to the range $[0, \pi]$, ensuring they correspond to valid rotation angles. The computation begins with the two-qubit system in the initial state $|00\rangle$. The input feature pair $(x_i, x_j)$ is loaded into the quantum state using the angle encoding method~\cite{schuld2021machine}, where the normalized feature values are used as rotation angles for $\boldsymbol{R}_y$ gates. This encoding step is represented by the unitary operator
\begin{equation}
	\label{eq:angle_encoding}
	\boldsymbol{U}_{\text{enc}}(x_i, x_j) = \boldsymbol{R}_y^{(0)}(x_i) \otimes \boldsymbol{R}_y^{(1)}(x_j),
\end{equation}
where $\boldsymbol{R}_y^{(k)}(x_k)$ denotes an $\boldsymbol{R}_y$ rotation by angle $x_k$ on the $k$-th qubit. The encoded state is then transformed by a trainable PQC $\boldsymbol{U}_q(\boldsymbol{\theta})$, where $\boldsymbol{\theta} \in \mathbb{R}^P$ is the vector of $P$ trainable parameters, yielding the final state as
\begin{equation}
	\label{eq:final_state}
	|\psi_{\text{final}}\rangle = \boldsymbol{U}_q(\boldsymbol{\theta}) \boldsymbol{U}_{\text{enc}}(x_i, x_j) |00\rangle.
\end{equation}

To investigate the impact of circuit architecture on performance, we designed and evaluated seven distinct trainable circuits, labeled A through G, as illustrated in Figure~\ref{fig:circuit}. These circuits, built from single-qubit rotation gates and CNOT gates, vary significantly in their parameter count, symmetry, and entanglement capability. 

To characterize these architectures, we quantified two key properties following established methods~\cite{sim2019expressibility} as
\begin{equation}
	\label{eq:expressibility}
	\text{Expr} = D_{\text{KL}}\left(\hat{P}_{\text{QNN}}(F; \boldsymbol{\theta}) \parallel P_{\text{Haar}}(F)\right),
\end{equation}
where $F$ denotes fidelity, $\boldsymbol{\theta}$ represents the trainable parameters, $\hat{P}_{\text{QNN}}$ is the state distribution generated by the quantum circuit, and $P_{\text{Haar}}$ is the Haar random state distribution. For consistency and comparability with existing PQC literature, we follow Sim et al.~\cite{sim2019expressibility} and QA-HFNN and adopt KL divergence as the expressibility metric. It is important to note that while this metric is commonly used as an indicator of state-space coverage capability, higher expressibility does not necessarily translate into improved classification performance. 

Second, the entangling capability was estimated using the Meyer-Wallach entanglement measure as
\begin{equation}
	\label{eq:entanglement}
	\text{Ent} = \frac{1}{|\mathcal{S}|} \sum_{\boldsymbol{\theta}_i \in \mathcal{S}} Q(|\psi_{\boldsymbol{\theta}_i}\rangle),
\end{equation}
where $Q$ represents the Meyer-Wallach entanglement measurement, $\mathcal{S}$ is the sampled parameter set, and $|\psi_{\boldsymbol{\theta}_i}\rangle$ denotes the quantum state generated by the circuit with parameter configuration $\boldsymbol{\theta}_i$. For two-qubit circuits, the Meyer-Wallach measure reduces to a normalized variant of linear entropy and yields consistent trends with established two-qubit entanglement measures such as concurrence. We therefore adopt it for consistency with prior QML literature. Higher values indicate stronger entanglement generation capability. 

These architectures span a controlled spectrum of expressivity and entanglement capacity, enabling a systematic study of how circuit structure affects relational learning. The detailed empirical analysis and performance comparison of these architectures are presented in Section~\ref{subsubsec:circuit_architecture_analysis}.

Following the evolution, we measure the expectation values of the Pauli-Z operator on each qubit ($\boldsymbol{M}_0 = \boldsymbol{Z} \otimes \boldsymbol{I}$ and $\boldsymbol{M}_1 = \boldsymbol{I} \otimes \boldsymbol{Z}$). The Pauli-Z expectation serves as a continuous observable whose range $[-1, 1]$ can be linearly mapped to fuzzy membership degrees. The two components of the joint membership vector are obtained by normalizing these expectation values to the range $[0, 1]$ as
\begin{equation}
	\label{eq:membership_0}
	\mu_0(x_i, x_j) = \frac{1}{2} \left(\langle\psi_{\text{final}}| \boldsymbol{M}_0 |\psi_{\text{final}}\rangle + 1\right),
\end{equation}
\begin{equation}
	\label{eq:membership_1}
	\mu_1(x_i, x_j) = \frac{1}{2} \left(\langle\psi_{\text{final}}| \boldsymbol{M}_1 |\psi_{\text{final}}\rangle + 1\right).
\end{equation}

Together, $\mu_0$ and $\mu_1$ form a two-dimensional fuzzy representation of the feature pair: $f_{\boldsymbol{\theta}}(x_i, x_j) = [\mu_0(x_i, x_j), \mu_1(x_i, x_j)]$.

\subsubsection{The Role of Entanglement in Relational Learning}
\label{subsubsec:entanglement_role}

Our trainable circuits $\boldsymbol{U}_q(\boldsymbol{\theta})$ are constructed from a universal gate set for two-qubit computation (single-qubit rotations and the CNOT gate). While this provides the circuits with high theoretical expressivity, our ablation experiments show that, under comparable expressivity and circuit depth, circuits capable of generating entanglement tend to yield better relational accuracy. This suggests that entanglement provides a beneficial structural property for modeling pairwise dependencies. By using the CNOT gate to create correlations between the two qubits, the model is empowered to directly learn the relationship between the paired features $(x_i, x_j)$, rather than treating them as independent pieces of information.

\subsubsection{Fuzzy Rule Layer}
\label{subsubsec:fuzzy_rule_layer}
For each of the $C$ classes, the model uses a dedicated dual-qubit circuit to process all $N_p$ sampled pairs. The resulting dual membership degrees $[\mu_{0,p}^{(c)}, \mu_{1,p}^{(c)}]$ are first combined via a product operation ($\mu_0 \cdot \mu_1$) to form a joint membership score for the pair. These scores are then aggregated by a fuzzy rule layer, which implements a conjunctive aggregation rule for evidence combination. To ensure numerical stability and feature normalization, this aggregation is performed using the Log-Geometric Mean operator, which retains the conjunctive semantics of the product t-norm while mitigating numerical underflow. For each class $c$, this process computes a final fuzzy feature score, $h_{\text{fuzzy}}^{(c)}$, as follows:
\begin{equation}
	\label{eq:fuzzy_aggregation}
	h_{\text{fuzzy}}^{(c)} = \frac{1}{N_p} \sum_{p=1}^{N_p} \log \left( \mu_{0,p}^{(c)} \cdot \mu_{1,p}^{(c)} \right),
\end{equation}
where $[\mu_{0,p}^{(c)}, \mu_{1,p}^{(c)}]$ is the joint membership vector for the $p$-th pair, specific to class $c$. This aggregation yields a $C$-dimensional vector $\boldsymbol{h}_{\text{fuzzy}}$ as the final output of the quantum branch.

\subsection{Classical DNN Branch}
\label{subsec:classical_dnn_branch}

The classical branch extracts high-dimensional semantic features using standard deep learning architectures. For the $l$-th layer, the feature vector is computed as
\begin{equation}
	\label{eq:classical_layer}
	\boldsymbol{h}^{(l)} = \sigma(\boldsymbol{W}^{(l)} \boldsymbol{h}^{(l-1)} + \boldsymbol{b}^{(l)}),
\end{equation}
where $\boldsymbol{W}^{(l)}$ and $\boldsymbol{b}^{(l)}$ are learnable parameters, and $\sigma$ is the activation function. Architecture specifications for each dataset are detailed in Table~\ref{tab:classical_architecture_CNN}.

\begin{table}[!t]
	\centering
	\caption{Classical Branch Architectures}
	\label{tab:classical_architecture_CNN}
	\renewcommand{\arraystretch}{1.2}
	\begin{small}
		\begin{tabular}{>{\centering\arraybackslash}p{2.2cm}>{\centering\arraybackslash}p{5.0cm}}
			\toprule[1pt]
			Dataset & Architecture \\
			\midrule[0.5pt]
			CIFAR-10 & ResNet-style (4 blocks, 64$\rightarrow$512 channels) \\
			Fashion-MNIST & 2-layer CNN + FC(320, 128) \\
			Dirty-MNIST & Same as Fashion-MNIST \\
			JAFFE & 6-layer CNN with residual blocks \\
			15-Scene & 3-layer MLP (200, 256) \\
			\bottomrule[1pt]
		\end{tabular}
	\end{small}
	\begin{tablenotes}
		\item \footnotesize All architectures use ReLU activation and batch normalization where applicable. CIFAR-10~\cite{krizhevsky2009learning} uses residual connections; Fashion-MNIST~\cite{xiao2017fashion}/Dirty-MNIST use dropout (0.3); 15-Scene~\cite{lazebnik2006beyond} uses dropout (0.4). FC denotes fully connected layers.
	\end{tablenotes}
\end{table}

\subsection{Pairing Strategy for Relational Feature Learning}
\label{subsec:hybrid_pairing_strategy}

A key innovation of our model is the hybrid pairing strategy, which selects feature pairs at multiple spatial scales by combining deterministic local pairing with stochastic global pairing. For image data, pixels are partitioned into a coarse grid; pairs are drawn both from within the same block (capturing local texture cues) and across distant blocks (capturing global context). For vector-based features (e.g., SIFT descriptors), proximity in index space replaces spatial distance. This flexibility allows the quantum branch to capture a diverse set of relational cues. 

\subsection{Fusion and Classifier Layers}
\label{subsec:fusion_classifier}

The features extracted from the parallel quantum and classical branches are integrated in the fusion layer. This process is based on the idea of multimodal fusion, which merges outputs from multiple networks to form a more comprehensive representation for the final classification task.

First, to ensure dimensional compatibility, the $C$-dimensional fuzzy feature vector $\boldsymbol{h}_{\text{fuzzy}}$ is projected into the $D$-dimensional classical feature space via a FC layer as
\begin{equation}
	\label{eq:fuzzy_projection}
	\boldsymbol{h}'_{\text{fuzzy}} = \boldsymbol{W}_f \boldsymbol{h}_{\text{fuzzy}} + \boldsymbol{b}_f,
\end{equation}
where $\boldsymbol{W}_f \in \mathbb{R}^{D \times C}$ and $\boldsymbol{b}_f \in \mathbb{R}^D$ are the learnable weights and bias of the fusion linear layer, respectively.

Next, the aligned fuzzy features $\boldsymbol{h}'_{\text{fuzzy}}$ and the classical features $\boldsymbol{h}_{\text{classical}}$ are fused using an additive fusion strategy. We adopt additive fusion for its parameter efficiency and ability to align fuzzy and semantic features in a shared embedding space, avoiding the dimensional inflation of concatenation-based fusion. The combined feature vector, $\boldsymbol{h}_{\text{fused}}$, is obtained by element-wise addition as
\begin{equation}
	\label{eq:fusion}
	\boldsymbol{h}_{\text{fused}} = \boldsymbol{h}_{\text{classical}} + \boldsymbol{h}'_{\text{fuzzy}}.
\end{equation}

Finally, the classifier layer processes the fused feature vector $\boldsymbol{h}_{\text{fused}}$ and classifies it into one of the $C$ categories. The classifier, implemented as a multilayer perceptron (MLP), maps the $D$-dimensional fused features to a $C$-dimensional output vector of logits, denoted as $\tilde{\boldsymbol{y}} = [\tilde{y}_1, \tilde{y}_2, \ldots, \tilde{y}_C]$. A softmax function is then utilized to convert these logits into a probability distribution over the classes, yielding the final predicted label probabilities as
\begin{equation}
	\label{eq:softmax}
	\hat{y}_c = \text{Softmax}(\tilde{\boldsymbol{y}})_c = \frac{e^{\tilde{y}_c}}{\sum_{k=1}^{C} e^{\tilde{y}_k}},
\end{equation}
where $\hat{y}_c$ is the predicted probability for the $c$-th class.

\subsection{Model Training}
\label{subsec:model_training}

\textbf{Loss Function:} We employ the categorical cross-entropy loss function. For a batch of $m$ training samples, the loss $L$ is calculated as
\begin{equation}
	\label{eq:loss_function}
	L = -\frac{1}{m} \sum_{i=1}^{m} \sum_{c=1}^{C} y_{i,c} \log(\hat{y}_{i,c}),
\end{equation}
where $y_{i,c}$ is a binary indicator (1 if the true label of sample $i$ is class $c$, and 0 otherwise), and $\hat{y}_{i,c}$ is the model's predicted probability for sample $i$ belonging to class $c$.

\textbf{Parameter Initialization:} For the quantum component, the trainable parameters $\boldsymbol{\theta}$ of the rotation gates are initialized by sampling from a uniform distribution over the interval $[0, 2\pi]$, i.e., $\boldsymbol{\theta} \sim U(0, 2\pi)$.

\textbf{Parameter Optimization:} The gradients for the classical components are calculated using the standard backpropagation algorithm. For the quantum membership function layer, the gradient can be obtained analytically using the parameter-shift rule~\cite{mitarai2018quantum,schuld2019evaluating}. The parameter-shift rule applies because all trainable gates in $\boldsymbol{U}_q(\boldsymbol{\theta})$ are single-parameter rotations of the form $\exp(-i\theta \hat{P}/2)$, where $\hat{P}$ is a Pauli operator. The gradient of a quantum circuit's expectation value output, $f_{\boldsymbol{\theta}}(\boldsymbol{x})$, with respect to parameter $\theta_i$ is given by
\begin{equation}
	\label{eq:parameter_shift}
	\frac{\partial f_{\boldsymbol{\theta}}(x)}{\partial \theta_i} = \frac{1}{2} \left( f_{\boldsymbol{\theta} + \frac{\pi}{2} \boldsymbol{e}_i}(x) - f_{\boldsymbol{\theta} - \frac{\pi}{2} \boldsymbol{e}_i}(x) \right),
\end{equation}
where $\boldsymbol{\theta} = [\theta_1, \theta_2, \ldots, \theta_P]$ is the vector of trainable parameters, and $\boldsymbol{\theta} \pm \frac{\pi}{2} \boldsymbol{e}_i$ represents the same vector but with only the $i$-th parameter shifted by $\pm\frac{\pi}{2}$. This enables exact, noise-free gradient computation without numerical finite-difference approximations.

The entire gradient computation process is integrated within the PyTorch framework, enabling end-to-end training. The complete training procedure is summarized in Algorithm~\ref{alg:dqhfnn_training}. The specific optimization strategy, including the choice of optimizer (e.g., AdamW~\cite{loshchilov2017decoupled} or SGD) and learning rate schedule, was tailored for each dataset. Detailed hyperparameters for each dataset are provided in Table~\ref{tab:hyperparameters}.

\begin{algorithm}[!t]
	\caption{DQ-HFNN Model Training Procedure}
	\label{alg:dqhfnn_training}
	\begin{algorithmic}[1]
		\REQUIRE Training dataset $\mathcal{D} = \{(\boldsymbol{X}^{(i)}, \boldsymbol{Y}^{(i)})\}$, learning rate $\alpha$, epochs $N$, pairs $N_{\text{p}}$
		\ENSURE Trained model parameters $\boldsymbol{\theta}$
		
		\STATE \textbf{Initialize Components:}
		\STATE \quad Quantum Layer: $\textsc{DualQubitFuzzy}(N_{\text{p}}, C)$
		\STATE \quad Pairing Strategy: Initialize $\textsc{Pairing}()$
		\STATE \quad Classical DNN: Initialize $\textsc{Dnn}()$
		\STATE \quad Fusion Layer: $\textsc{Linear}(C, D)$
		\STATE \quad Classifier: $\textsc{Classifier}(D, C)$
		
		\STATE \textbf{Training Loop:}
		\FOR{$e = 1, \ldots, N$}
		\FOR{each batch $(\boldsymbol{X}, \boldsymbol{Y})$}
		\STATE $\mathcal{P} \leftarrow \textsc{Pairing}(\boldsymbol{X})$ 
		\STATE $\boldsymbol{H}_{\text{fuz}} \leftarrow \textsc{DualQubitFuzzy}(\mathcal{P})$
		\STATE $\boldsymbol{H}_{\text{cnn}} \leftarrow \textsc{Dnn}(\boldsymbol{X})$
		\STATE $\boldsymbol{H}_{\text{rule}} \leftarrow \mathrm{PROD}(\boldsymbol{H}_{\text{fuz}}, \text{dim}=1)$
		\STATE $\boldsymbol{H}_{\text{fused}} \leftarrow \mathrm{ADD}(\textsc{Linear}(\boldsymbol{H}_{\text{rule}}), \boldsymbol{H}_{\text{cnn}})$
		\STATE $\hat{\boldsymbol{Y}} \leftarrow \mathrm{Softmax}(\textsc{Classifier}(\boldsymbol{H}_{\text{fused}}))$
		\STATE Compute loss $L$ using Eq.~(\ref{eq:loss_function}) with $\hat{\boldsymbol{Y}}$ and $\boldsymbol{Y}$
		\STATE Compute gradients via backpropagation (classical) 
		\STATE \quad and parameter-shift rule (quantum)
		\STATE Update all parameters $\boldsymbol{\theta}$ using optimizer
		\ENDFOR
		\ENDFOR
	\end{algorithmic}
\end{algorithm}
\section{EXPERIMENTS}
\label{sec:experiments}

\subsection{Experimental Setup}
\label{subsec:experimental_setup}

\subsubsection{Environment and Implementation}

All simulations were conducted on a single NVIDIA RTX 4060 GPU. The classical deep learning components were implemented in PyTorch, while quantum circuits were simulated using the TorchQuantum~\cite{wang2022torchquantum} library, which provides seamless integration and automatic differentiation. Quantum circuits were simulated in a noiseless setting unless otherwise specified.

\subsubsection{Datasets and Preprocessing}

We evaluate DQ-HFNN on five heterogeneous benchmarks. For the JAFFE dataset, facial regions were extracted using a Haar Cascade face detector (no identity-specific tuning) and resized to $32 \times 32$ pixels. The Dirty-MNIST dataset is generated by the \texttt{ddu\_dirty\_mnist} library~\cite{mukhoti2023deep}, which inherently includes Gaussian noise ($\sigma = 0.05$) as part of its construction. The 15-Scene dataset is represented using 200-dimensional SIFT Bag-of-Words vectors. CIFAR-10 and Fashion-MNIST are used in their standard splits. Standard data augmentation (random cropping and horizontal flipping) was applied to CIFAR-10 and JAFFE during training. Fashion-MNIST and Dirty-MNIST used normalization only. All datasets were normalized to zero mean and unit variance using dataset-specific statistics.

\subsubsection{Model Configuration and Hyperparameters}

Our model employs an adaptive paired sampling strategy (detailed in Section~\ref{subsec:hybrid_pairing_strategy}), where the number of sampled pairs is determined as a fixed proportion of the maximum pairing budget. The 30\% sampling ratio, yielding 153 pairs for $32 \times 32$ images, is identified through ablation studies (Section~\ref{subsec:ablation_studies}) as optimal for balancing information capture and computational efficiency.

Training protocols were adapted based on dataset scale. JAFFE uses 8-fold cross-validation due to its small size (213 images total). Larger datasets employ conventional train-validation-test splits with standard training protocols. For JAFFE, we used the AdamW optimizer with a 3-epoch warm-up followed by CosineAnnealingLR. For other datasets, we used SGD with momentum 0.9 and MultiStepLR decay at epochs 56 and 78. All JAFFE and 15-Scene results are averaged over 40 independent runs to ensure statistical robustness. Key hyperparameters are summarized in Table~\ref{tab:hyperparameters}.

\begin{table}[!t]
	\centering
	\caption{Key Hyperparameter Settings for DQ-HFNN}
	\label{tab:hyperparameters}
	\renewcommand{\arraystretch}{1.3}
	\begin{small}
		\begin{tabular}{p{2.8cm}p{1.5cm}p{2cm}}
			\toprule[1pt]
			Hyperparameter & JAFFE & Other Datasets\\
			\midrule[0.5pt]
			Training Scheme & 8-Fold CV & Standard Split \\
			Batch Size & 8 & 128 \\
			Initial LR & 5e-4 & 1e-2 \\
			Optimizer & AdamW & SGD \\
			Activation & ReLU & ReLU \\
			Measurement & Pauli-Z & Pauli-Z \\
			Encoding Gate & $R_y$ & $R_y$ \\
			\thead[l]{Random Pairs \\ (30\% sampling)} & \thead[c]{153 pairs \\ for $32 \times 32$} & \thead[c]{$\frac{\mathrm{n}^2}{2} \times 30~\%$ \\ for $\mathrm{n}\times \mathrm{n}$} \\
			\bottomrule[1pt]
		\end{tabular}
	\end{small}
	\begin{tablenotes}
		\item \footnotesize ``Other Datasets'' includes CIFAR-10, Fashion-MNIST, Dirty-MNIST, and 15-Scene. $\mathrm{n}$ denotes the side length of the input image.CV denotes Cross-Validation train scheme.
	\end{tablenotes}
\end{table}

\subsubsection{Evaluation Metrics and Protocol}

Model performance was assessed using Accuracy, macro-averaged Precision, Recall, and F1-Score. Macro-averaged metrics are used because some datasets (e.g., JAFFE) exhibit class imbalance, and macro averaging treats each class equally. To ensure statistical significance, particularly for smaller datasets, all reported results for JAFFE and 15-Scene are the mean and standard deviation over 40 independent runs. Each run uses a different random seed for data shuffling, initialization, and pair sampling to ensure statistical robustness.

\subsection{Performance Comparison on Benchmark Datasets}
\label{subsec:performance_comparison}

\subsubsection{Overall Performance on General Datasets}

To evaluate the general effectiveness and versatility of our proposed DQ-HFNN model, we first compare its performance against several baselines on four diverse datasets. The results are summarized in Table~\ref{tab:overall_performance}.

\begin{table*}[!t]
	\centering
	\begin{threeparttable}
		\caption{Accuracy Comparison on General Benchmark Datasets}
		\label{tab:overall_performance}
		\begin{tabular}{>{\centering\arraybackslash}p{3.2cm}>{\centering\arraybackslash}p{3.2cm}>{\centering\arraybackslash}p{3.2cm}>{\centering\arraybackslash}p{3.2cm}>{\centering\arraybackslash}p{3.2cm}}
			\toprule[1pt]
			Model & CIFAR-10 & JAFFE & Fashion-MNIST & 15-Scene \\
			\midrule[0.5pt]
			DNN & 0.9427 & 0.9400 $\pm$ 0.0422 & 0.9072 & 0.7232 $\pm$ 0.0050 \\
			FDNN & 0.9436 & 0.9353 $\pm$ 0.0465 & 0.9075 & 0.7225 $\pm$ 0.0048 \\
			QA-HFNN & 0.9430 & 0.9446 $\pm$ 0.0400 & 0.9080 & 0.7244 $\pm$ 0.0046 \\
			DQ-HFNN & \textbf{0.9446} & \textbf{0.9495 $\pm$ 0.0379} & \textbf{0.9096} & \textbf{0.7252 $\pm$ 0.0065} \\
			\bottomrule[1pt]
		\end{tabular}
		\begin{tablenotes}
			\item \footnotesize The bold values represent the metrics of our proposed model. Results with standard deviations (JAFFE and 15-Scene) are averaged over 40 independent runs. DQ-HFNN uses Circuit C architecture, identified as optimal in Section~\ref{subsec:ablation_studies}.
		\end{tablenotes}
	\end{threeparttable}
\end{table*}

As shown in Table~\ref{tab:overall_performance}, DQ-HFNN achieves consistent, though modest, improvements over both classical and single-qubit baselines. On the JAFFE dataset, DQ-HFNN exceeds QA-HFNN by 0.49 percentage points on average; given the variance of this dataset, this suggests a stable performance gain. Similar improvements are observed on the other datasets. The slightly lower performance of FDNN on JAFFE is likely due to its larger parameter count, which makes it more prone to overfitting on small datasets.

\subsubsection{In-depth Analysis on the Noisy Dirty-MNIST Dataset}

To further assess the model's robustness under significant data uncertainty, we conducted a detailed analysis on the Dirty-MNIST dataset. The comprehensive performance metrics are presented in Table~\ref{tab:dirty_mnist_performance}. For this dataset, the primary result for our model is reported using the non-entangled Circuit A (DQ-HFNN-A), because Circuit A achieves the best performance.

\begin{table}[!t]
	\centering
	\begin{threeparttable}
		\caption{Performance Analysis on the Dirty-MNIST Dataset}
		\label{tab:dirty_mnist_performance}
		\renewcommand{\arraystretch}{1.3}
		\begin{small}
			\begin{tabular}{>{\centering\arraybackslash}p{1.8cm}>{\centering\arraybackslash}p{1.175cm}>{\centering\arraybackslash}p{1.175cm}>{\centering\arraybackslash}p{1.175cm}>{\centering\arraybackslash}p{1.175cm}}
				\toprule[1pt]
				Model & Accuracy & Precision & Recall & F1 Score \\
				\midrule[0.5pt]
				FDNN & 0.839 & 0.856 & 0.849 & 0.852 \\
				QA-HFNN & 0.840 & 0.858 & 0.849 & 0.853 \\
				ResNet18 & 0.836 & 0.857 & 0.850 & 0.852 \\
				ResNet50 & 0.835 & 0.851 & 0.845 & 0.848 \\
				DQ-HFNN-A & \textbf{0.840} & \textbf{0.859} & \textbf{0.850} & \textbf{0.854} \\
				DQ-HFNN-B & 0.839 & 0.855 & 0.848 & 0.851 \\
				DQ-HFNN-C & 0.839 & 0.857 & 0.849 & 0.852 \\
				\bottomrule[1pt]
			\end{tabular}
		\end{small}
		\begin{tablenotes}
			\item \footnotesize The bold values represent the metrics of our proposed model with the best-performing circuit configuration.
		\end{tablenotes}
	\end{threeparttable}
\end{table}

On Dirty-MNIST, the non-entangled variant DQ-HFNN-A achieves marginally higher accuracy and F1-score. Given the extremely small differences ($\leq$ 0.1\%), we refrain from attributing this to a definitive architectural cause. One possible explanation is that simpler circuits may generalize slightly better under heavy noise, consistent with the observation that ResNet18 outperforms the deeper ResNet50.

\subsubsection{Training Stability and Convergence}

\begin{figure}[!t]
	\centering
	\includegraphics[scale=0.95]{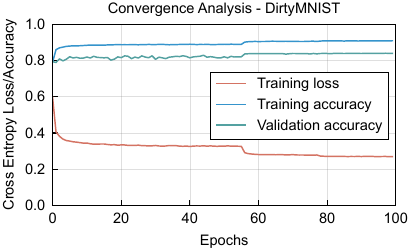}
	\caption{Training and validation curves for the DQ-HFNN model on the Dirty-MNIST dataset. The training loss decreases smoothly, with a scheduled drop at epoch 56 corresponding to learning rate decay. Both training and validation accuracies rise steadily and reach a stable plateau, demonstrating effective optimization.}
	\label{fig:convergence_analysis}
\end{figure}

To validate the training stability of our proposed model, a representative convergence analysis on the Dirty-MNIST dataset is presented in Figure~\ref{fig:convergence_analysis}.

Figure~\ref{fig:convergence_analysis} shows the convergence behavior of DQ-HFNN on Dirty-MNIST, suggesting a well-behaved optimization process.
\subsection{Ablation Studies}
\label{subsec:ablation_studies}
To deconstruct the sources of performance improvement in our DQ-HFNN model, we conducted a series of ablation studies. This section analyzes the quantum pairing and sampling strategy. All experiments in this section were performed on the JAFFE dataset unless otherwise specified. Mutual information was estimated using the k-nearest neighbor (kNN) estimator with $k = 3$~\cite{kraskov2004estimating}, as implemented in scikit-learn's \texttt{mutual\_info\_classif} function.

\subsubsection{Analysis of the Quantum Pairing and Sampling Strategy}

\paragraph{Impact of Pairing Configuration}

To isolate the effects of the pairing strategy from entanglement, these experiments utilized a specially designed non-entangled dual-qubit circuit that maintains the same gate structure and parameter count as Circuit C. We evaluated four distinct pairing configurations, labeled `a' through `d', to understand the trade-off between fixed local pairs and fully random pairs. The configurations are: (a) 512 pairs with a 30\% random and 70\% fixed split; (b) 512 pairs with a 50\% random and 50\% fixed split; (c) 153 fully random pairs; and (d) 256 fully random pairs. Configurations (c) and (d) use 153 and 256 fully random pairs, respectively, matching the random pair counts in hybrid configurations (a) and (b). This design enables controlled comparison between pure random and hybrid strategies under equivalent sampling budgets.

\begin{table}[!t]
	\centering
	\begin{threeparttable}
		\caption{Validation Accuracy of Different Pairing Strategies on JAFFE}
		\label{tab:pairing_strategies}
		\renewcommand{\arraystretch}{1.3}
		\begin{small}
			\begin{tabular}{>{\centering\arraybackslash}p{4.0cm}>{\centering\arraybackslash}p{3.5cm}}
				\toprule[1pt]
				Model & Accuracy \\
				\midrule[0.5pt]
				512-30\%(a) & 0.944 $\pm$ 0.035 \\
				512-50\%(b) & 0.940 $\pm$ 0.043 \\
				153-100\%(c) & \textbf{0.948 $\pm$ 0.039} \\
				256-100\%(d) & 0.935 $\pm$ 0.046 \\
				\bottomrule[1pt]
			\end{tabular}
		\end{small}
		\begin{tablenotes}
			\item \footnotesize The bold values represent the best-performing configuration.
		\end{tablenotes}
	\end{threeparttable}
\end{table}

The validation accuracies, reported in Table~\ref{tab:pairing_strategies}, indicate that configuration (c) achieves the best mean accuracy under our experimental setting. However, given the overlapping standard deviations, the improvement over configurations (a) and (b) should be interpreted as a stable trend rather than a statistically significant difference. To elucidate the underlying factors, we conducted a comprehensive information-theoretic analysis, using the following formulation
\begin{equation}
	\text{Red}(F_q, F_c; L) = \mathrm{I}(F_q; L) + \mathrm{I}(F_c; L) - \mathrm{I}(F_q, F_c; L),
\end{equation}
where $\mathrm{I}(\cdot;\cdot)$ denotes mutual information. Positive values indicate redundancy (information overlap), while negative values indicate synergy. Lower positive values suggest more complementary representations. Therefore, we can quantify the information relationship between the quantum ($F_q$) and classical ($F_c$) feature sets with respect to the labels ($L$).

\begin{figure*}[!t]
	\centering
	\includegraphics[scale=0.95]{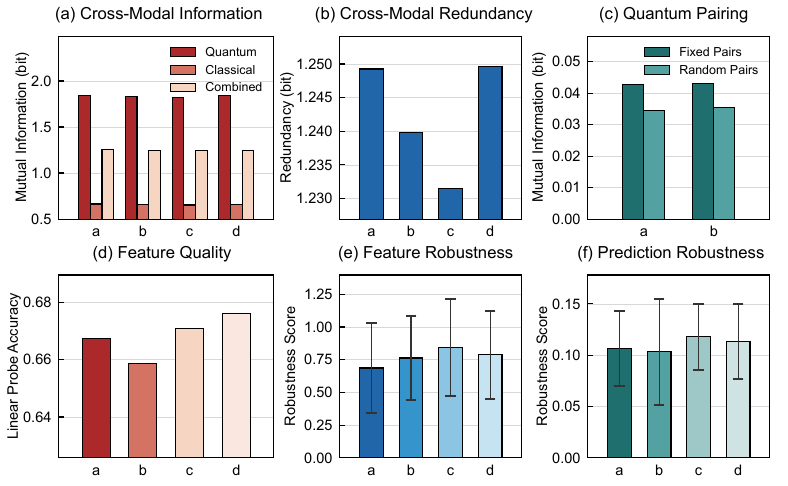}
	\caption{Multi-dimensional analysis of four distinct pairing strategies (a--d) on the JAFFE dataset. (a) Mutual information between features (Quantum, Classical, Combined) and labels. (b) Information redundancy between quantum and classical branches. Lower values indicate more complementary feature learning.(c) Information contribution of fixed vs. random pairs. (d) Fused feature quality measured by Linear Probe Accuracy. (e, f) Robustness of features and predictions against input noise (Gaussian noise with $\sigma = 0.05$). The robustness scores are computed by subtracting the raw distance/divergence metrics from their maximum values across all configurations, such that higher scores indicate better robustness. The best-performing strategy (c) (153 random pairs) demonstrates a favorable balance of information quality and prediction robustness.}
	\label{fig:combined_analysis}
\end{figure*}

Figure~\ref{fig:combined_analysis}(a) shows that the quantum component consistently contributes substantial information about the labels. Figure~\ref{fig:combined_analysis}(b) shows that all configurations exhibit positive redundancy values, where configuration (c) achieves the lowest redundancy, indicating the most complementary information distribution between the two branches. Figure~\ref{fig:combined_analysis}(c) reveals that in hybrid strategies, the information contribution from fixed pairs is similar to that from random pairs, suggesting potential redundancy when using a large number of fixed pairs.

Feature Quality, assessed by Linear Probe Accuracy in Figure~\ref{fig:combined_analysis}(d), measures the linear separability of the fused features. Robustness against input noise is quantified by the Euclidean distance (for features) and KL divergence (for predictions) between original and noisy inputs, where Gaussian noise with $\sigma = 0.05$ was added to the inputs. The displayed robustness scores are computed by subtracting the raw metrics from their maximum values across all configurations, such that higher scores indicate better robustness. As shown in Figure~\ref{fig:combined_analysis}(e) and (f), configuration (c) achieves competitive prediction robustness. The observation that configuration (d) has high feature quality but lower accuracy may indicate that it produces more separable features in a linear space but not necessarily features that generalize well, possibly due to sensitivity to noise.

\paragraph{Sparse Sampling as an Implicit Regularization Mechanism}

\begin{figure*}[!t]
	\centering
	\includegraphics[scale=0.95]{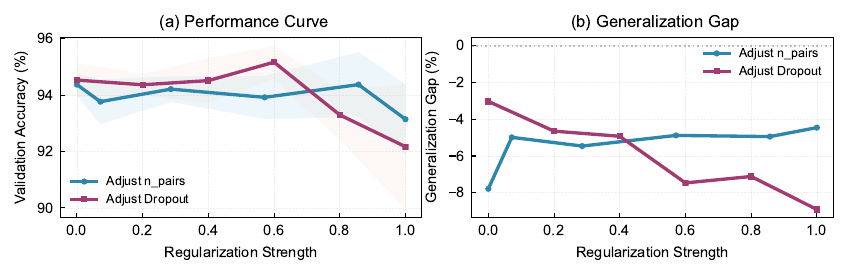}
	\caption{Comparison of sparse sampling and Dropout as regularization mechanisms. (a) Validation accuracy as a function of normalized regularization strength. Adjusting the number of pairs exhibits a broader optimal range than adjusting the Dropout rate. (b) Both methods effectively control the generalization gap (training minus validation accuracy), but their distinct curve shapes suggest they influence the model in qualitatively different ways.}
	\label{fig:regularization_comparison}
\end{figure*}

To explore whether the number of sampled pairs acts as an implicit regularizer, we compared the effect of adjusting the number of random pairs against that of varying the Dropout rate, a standard regularization technique. The results are shown in Figure~\ref{fig:regularization_comparison}.

Figure~\ref{fig:regularization_comparison}(a) plots validation accuracy against a normalized ``Regularization Strength''. The curve for adjusting pairs (blue line) shows a broad performance plateau, indicating robustness to this hyperparameter, while the Dropout curve (purple line) displays a more sensitive peak. Figure~\ref{fig:regularization_comparison}(b) shows that both methods control the generalization gap (training minus validation accuracy). However, their distinct curve shapes, confirmed by a low correlation between their second derivatives ($r = -0.0032$), suggest that the two regularization mechanisms influence the model in qualitatively different ways, although a deeper theoretical analysis is required to fully characterize this distinction.

This analysis suggests that reducing the number of random pairs may serve as an intrinsic regularization mechanism, distinct from Dropout. The favorable performance of configuration (c) is thus attributed to a balance of information extraction capacity and this implicit regularization effect.

\subsubsection{Analysis of the Quantum Circuit Architecture}
\label{subsubsec:circuit_architecture_analysis}

As described in Section~\ref{subsubsec:quantum_implementation}, we designed seven circuit architectures (A--G) with varying expressibility and entanglement properties. We evaluated six of the seven architectures, excluding the highly complex Circuit F due to unstable optimization behavior observed in preliminary experiments. The theoretical metrics (expressibility and entanglement) are computed alongside empirical classification accuracy on the JAFFE dataset. The consolidated results are presented in Table~\ref{tab:jaffe_circuits}.

\begin{table}[!t]
	\centering
	\begin{threeparttable}
		\caption{Theoretical Properties and Performance of Circuit Architectures on JAFFE}
		\label{tab:jaffe_circuits}
		\renewcommand{\arraystretch}{1.3}
		\begin{small}
			\begin{tabular}{>{\centering\arraybackslash}p{1.0cm}>{\centering\arraybackslash}p{1.8cm}>{\centering\arraybackslash}p{1.8cm}>{\centering\arraybackslash}p{1.9cm}}
				\toprule[1pt]
				Arch & Expressibility & Entanglement & Accuracy \\
				\midrule[0.5pt]
				A & 0.940 & 0.000 & 0.945 $\pm$ 0.040 \\
				B & 0.525 & 0.302 & 0.949 $\pm$ 0.039 \\
				C & 0.531 & 0.476 & 0.950 $\pm$ 0.038 \\
				D & 0.003 & 0.535 & 0.936 $\pm$ 0.044 \\
				E & 0.044 & 0.000 & 0.940 $\pm$ 0.040 \\
				F & 0.002 & 0.575 & --- \\
				G & 0.199 & 0.000 & 0.942 $\pm$ 0.043 \\
				\bottomrule[1pt]
			\end{tabular}
		\end{small}
		\begin{tablenotes}
			\item \footnotesize Lower expressibility metric values indicate higher expressibility capability (i.e., closer to the Haar distribution).
		\end{tablenotes}
	\end{threeparttable}
\end{table}

The results in Table~\ref{tab:jaffe_circuits} show that the entangled asymmetric Circuits C and B achieve the highest mean accuracies (0.950 and 0.949, respectively). Notably, Circuits D and E exhibit high expressibility (KL divergence close to 0) but yield lower accuracies (0.936 and 0.940), while Circuits C and B with moderate expressibility but strong entanglement (0.476 and 0.302) outperform them. This suggests that relational modeling enabled by entanglement plays a more decisive role than state-space coverage capability alone.

To validate these findings, we extended the comparison of the top-performing Circuits (A, B, C) to the remaining datasets. The results are presented in Table~\ref{tab:circuits_all_datasets}. Although the mean accuracies follow the trend $C \geq B \geq A$, the differences on larger datasets (e.g., CIFAR-10 and Fashion-MNIST) are extremely small and should be interpreted as suggestive rather than conclusive. The notable exception is the Dirty-MNIST dataset (Table~\ref{tab:dirty_mnist_performance}), where the non-entangled Circuit A performs best, suggesting that in highly noisy environments, the relational modeling induced by entanglement might inadvertently capture spurious correlations.

\begin{table}[!t]
	\centering
	\begin{threeparttable}
		\caption{Extended Performance Comparison of Circuit Architectures A, B, and C}
		\label{tab:circuits_all_datasets}
		\begin{tabular}{>{\centering\arraybackslash}p{0.8cm}>{\centering\arraybackslash}p{1.0cm}>{\centering\arraybackslash}p{2.5cm}>{\centering\arraybackslash}p{2.2cm}}
			\toprule[1pt]
			Model & CIFAR-10 & Fashion-MNIST & 15-Scene \\
			\midrule[0.5pt]
			A & 0.944 & 0.907 $\pm$ 0.002 & 0.723 $\pm$ 0.006 \\
			B & 0.945 & 0.907 $\pm$ 0.001 & 0.723 $\pm$ 0.005 \\
			C & 0.945 & 0.908 $\pm$ 0.001 & 0.725 $\pm$ 0.007 \\
			\bottomrule[1pt]
		\end{tabular}
		\begin{tablenotes}
			\item \footnotesize Results for JAFFE and Dirty-MNIST are presented in Tables~\ref{tab:jaffe_circuits} and~\ref{tab:dirty_mnist_performance}, respectively.
		\end{tablenotes}
	\end{threeparttable}
\end{table}

\begin{figure}[!t]
	\centering
	\includegraphics[scale=0.95]{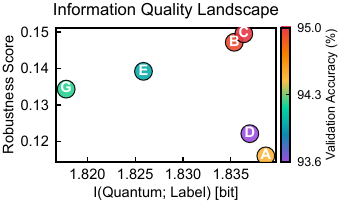}
	\caption{Information Quality Landscape of six circuit architectures (A--G, excluding F) on the JAFFE dataset. The x-axis represents the mutual information between the quantum features and the labels, indicating information relevance. The y-axis represents the model's prediction robustness score (computed as the maximum KL divergence minus the circuit-specific KL divergence, such that higher values indicate better robustness). The color of each point corresponds to its validation accuracy. Circuits C and B are located in a region that exhibits both high robustness and substantial information relevance.}
	\label{fig:info_landscape}
\end{figure}

Entanglement appears to play a key role in relational modeling. The consistent correlation between entanglement capacity and performance led us to investigate their underlying relationship. The Information Quality Landscape in Figure~\ref{fig:info_landscape} plots the relationship among model performance, robustness, and the mutual information between their quantum features and labels. Here, Circuits C and B, as the two best-performing architectures, are situated in a region that achieves a balance between high robustness and high information relevance. In contrast, Circuits A, D, and G, despite excelling in individual metrics, do not demonstrate comparable overall performance.

To explore whether entanglement enables learning of relational patterns, we conducted a structured perturbation experiment comparing the best entangled model (Circuit C) with its non-entangled counterpart (Circuit A). We applied two categories of perturbations: relation-preserving transforms, including Brightness ($\delta = 0.1$) and Contrast ($\alpha = 1.2$) adjustments, which shift absolute pixel values while maintaining relative differences; and relation-breaking transforms, including Local Shuffle ($3 \times 3$ window) and Global Shuffle, which disrupt spatial dependencies. To quantify the robustness of each circuit against these perturbations, we measured the KL divergence between the model's predictions on the original and perturbed inputs. A lower KL divergence indicates higher robustness (invariance) to the specific perturbation.

\begin{figure}[!t]
	\centering
	\includegraphics[scale=0.95]{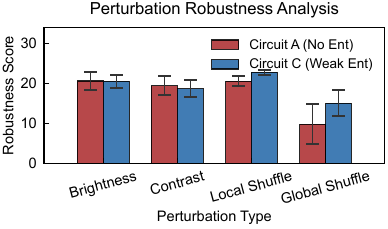}
	\caption{Perturbation robustness comparison between non-entangled (Circuit A) and entangled (Circuit C) architectures on the JAFFE dataset. The robustness score is derived from the KL divergence between predictions on original and perturbed inputs, where a lower KL divergence is visualized as a higher robustness score, indicating greater robustness. Circuit C demonstrates superior robustness under relation-breaking perturbations (Local Shuffle and Global Shuffle), suggesting that entanglement may enable the learning of more abstract, position-invariant relational patterns.}
	\label{fig:perturbation_robustness}
\end{figure}

The results, shown in Figure~\ref{fig:perturbation_robustness}, suggest a distinction between the two circuit types. Under relation-breaking perturbations (Shuffle), the entangled Circuit C demonstrates higher robustness than Circuit A, suggesting it may learn more abstract, position-invariant patterns. Conversely, under relation-preserving perturbations, both circuits exhibit similar robustness, indicating the downstream classical network also contributes to learning certain invariances.

In summary, our analysis suggests that the performance benefit associated with entanglement may stem from its ability to facilitate relational modeling rather than from increased expressibility alone. This relational capacity appears to enable learning of more abstract and robust feature representations.
\subsection{Further Analysis and Discussion}
\label{subsec:further_analysis}

\subsubsection{Robustness to Quantum Noise}
\label{subsubsec:quantum_noise_robustness}

To assess the feasibility of our proposed DQ-HFNN on near-term quantum devices, we analyzed the robustness of its dual-qubit circuits under simulated quantum noise. We evaluated four standard noise models: amplitude damping (AD), depolarizing (DP), bit flip (BF), and phase flip (PF)~\cite{nielsen2010quantum}.

The noisy quantum state is obtained via the operator-sum representation as
\begin{equation}
	\label{eq:operator_sum}
	\varepsilon(\boldsymbol{\rho}) = \sum_{k} \boldsymbol{E}_k \boldsymbol{\rho} \boldsymbol{E}_k^{\dagger},
\end{equation}
where $\{\boldsymbol{E}_k\}$ are the Kraus operators satisfying $\sum_k \boldsymbol{E}_k^{\dagger} \boldsymbol{E}_k = \boldsymbol{I}$.

To quantify the circuit's noise tolerance, we compute the fidelity between the ideal pure state $|\psi\rangle$ and the noisy density matrix $\boldsymbol{\rho}_{\text{noisy}}$. Since the target state is pure, the fidelity simplifies to the overlap as
\begin{equation}
	\label{eq:fidelity}
	\mathrm{F}(|\psi\rangle, \boldsymbol{\rho}_{\text{noisy}}) = \langle \psi | \boldsymbol{\rho}_{\text{noisy}} | \psi \rangle.
\end{equation}

For our evaluation, we sampled 200 input values uniformly in $[-1, 1]$ and varied the noise probability $\gamma$ from 0.01 to 0.1. The fidelity is averaged over all inputs for each noise level. The average fidelities for the CIFAR-10 task are summarized in Table~\ref{tab:noise_robustness}. 

\begin{table}[!t]
	\centering
	\begin{threeparttable}
		\caption{Average Fidelity of the Dual-Qubit Circuits on CIFAR-10 Under Various Noise Models}
		\label{tab:noise_robustness}
		\begin{tabular}{>{\centering\arraybackslash}p{1.2cm}>{\centering\arraybackslash}p{1.2cm}>{\centering\arraybackslash}p{1.2cm}>{\centering\arraybackslash}p{1.2cm}>{\centering\arraybackslash}p{1.2cm}}
			\toprule[1pt]
			$\gamma$ & AD & DP & BF & PF \\
			\midrule[0.5pt]
			0.01 & 0.9988 & 0.9920 & 0.9830 & 0.9931 \\
			0.03 & 0.9966 & 0.9768 & 0.9513 & 0.9801 \\
			0.05 & 0.9944 & 0.9617 & 0.9200 & 0.9675 \\
			0.07 & 0.9917 & 0.9427 & 0.8818 & 0.9522 \\
			0.10 & 0.9885 & 0.9200 & 0.8370 & 0.9344 \\
			\bottomrule[1pt]
		\end{tabular}
	\end{threeparttable}
\end{table}

The circuits maintain high fidelity across all tested noise levels and models. These results suggest that the shallow depth and limited entangling operations of our dual-qubit circuits contribute to their robustness on NISQ devices, as noise accumulation remains minimal.

\subsubsection{Gradient Stability}
\label{subsubsec:gradient_stability}
\begin{figure}[!t]
	\centering
	\includegraphics[scale=0.95]{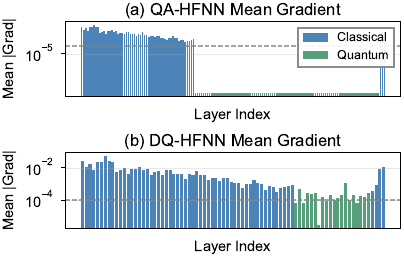}
	\caption{Comparison of gradient flows on the CIFAR-10. The blue and green bars represent the gradient flows of the classical and quantum branches, respectively. (a) The gradient flow in the quantum layer of QA-HFNN vanishes due to the product operation over 3072 dimensions. (b) In contrast, DQ-HFNN operates in the logarithmic domain, guaranteeing gradient stability and ensuring a stable training process.}
	\label{fig:stability}
\end{figure}

In contrast to the AND aggregation logic employed by QA-HFNN, we adopt a compensatory operator with AND-like conjunctive semantics, which stabilizes the gradient flow during training. As illustrated in Figure~\ref{fig:stability}, on the CIFAR-10, the gradient flow in the quantum layer of QA-HFNN vanishes, indicating that the quantum layer makes effectively no contribution during the training process. However, our DQ-HFNN remains highly stable. This effectively prevents numerical underflow and ensures gradient stability throughout the training process.

\subsubsection{Sampling Efficiency and Computational Complexity}
\label{subsubsec:sampling_efficiency_complexity}

\begin{figure}[!t]
	\centering
	\includegraphics[scale=0.95]{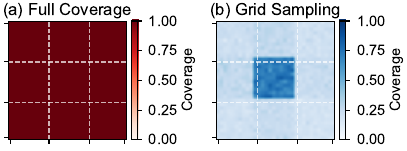}
	\caption{Comparison of sampling coverage strategies. (a) The full-coverage approach of QA-HFNN processes every input pixel. (b) Our proposed grid-based paired sampling for DQ-HFNN strategically covers the input space while significantly reducing the feature dimensionality, focusing sampling density on the central region.}
	\label{fig:sampling_heatmap}
\end{figure}

A key innovation of our DQ-HFNN is the paired sampling strategy, which contrasts with the full-coverage approach of single-qubit models like QA-HFNN. As visualized in Figure~\ref{fig:sampling_heatmap}, our grid-based sampling reduces the quantum feature dimensionality from 3072 (all pixels in a $32 \times 32 \times 3$ CIFAR-10 image) to approximately 918 (153 pairs $\times$ 2 qubits $\times$ 3 channels).

This reduction in dimensionality leads to computational efficiency. The theoretical time complexity of the quantum branch is $O(N_{\mathrm{p}})$, where $N_{\mathrm{p}}$ is the number of sampled pairs. Since $N_{\mathrm{p}}$ is a design parameter chosen to be smaller than the total number of input features $P$ (i.e., $N_{\mathrm{p}} \ll P$), this provides an asymptotic advantage over full-coverage methods. In terms of parameter efficiency, the quantum component of DQ-HFNN for a 10-class problem utilizes 20 trainable parameters, compared to 61,440 for FDNN and 90 for QA-HFNN. Regarding empirical inference time, DQ-HFNN is approximately 2.4 times faster than QA-HFNN on the CIFAR-10 benchmark, thereby optimizing the trade-off between relational modeling capability and computational cost.
\section{Conclusions}
\label{sec:conclusions}
In conclusion, this paper proposes a hybrid quantum-classical architecture DQ-HFNN for data uncertainty and relationship modeling in fuzzy systems. By extending the single-qubit system to a dual-qubit system, we demonstrate that quantum entanglement can be used to learn the joint membership function of feature pairs. Compared with the QA-HFNN, this approach enables the model to capture the correlation patterns and contextual relationships between features. The experimental results validate the effectiveness of the proposed DQ-HFNN model. Specifically, the DQ-HFNN demonstrates higher classification accuracy than classical benchmarks and QA-HFNN. The ablation studies show that the performance gain is mainly related to the relational modeling capabilities conferred by entanglement, rather than the increase in model expressivity. It is worth noting that further statistical studies are required to fully establish the contribution of entanglement. This model exhibits high parameter efficiency and robustness against simulated quantum noise, demonstrating its potential for practical implementation on NISQ devices.

Future work includes two main directions. First, the proposed DQ-HFNN model can be validated on larger-scale or higher-uncertainty datasets, especially in medical imaging applications such as chest X-rays and histopathology images. In addition, by incorporating attention mechanisms, the pairing strategy may be enhanced to select more meaningful feature pairs dynamically. These efforts aim to develop practical quantum machine learning algorithms for the NISQ era, preparing for powerful applications in the future fault-tolerant quantum computing regime.

%\bibliography{cite}

\begin{thebibliography}{99}
	
	\bibitem{zadeh1965fuzzy} L. A. Zadeh, ``Fuzzy sets,'' \textit{Information and Control}, vol. 8, no. 3, pp. 338--353, 1965.
	
	\bibitem{deng2016hierarchical} Y. Deng, Z. Ren, Y. Kong, F. Bao, and Q. Dai, ``A hierarchical fused fuzzy deep neural network for data classification,'' \textit{IEEE Transactions on Fuzzy Systems}, vol. 25, no. 4, pp. 1006--1012, 2016.
	
	\bibitem{roy2024fa} A. Roy, A. Bhattacharjee, D. Oliva, O. Ramos-Soto, F. J. Alvarez-Padilla, and R. Sarkar, ``FA-Net: A fuzzy attention-aided deep neural network for pneumonia detection in chest X-Rays,'' in \textit{2024 IEEE 37th International Symposium on Computer-Based Medical Systems (CBMS)}, 2024, pp. 338--343.
	
	\bibitem{yazdinejad2023optimized} A. Yazdinejad, A. Dehghantanha, R. M. Parizi, and G. Epiphaniou, ``An optimized fuzzy deep learning model for data classification based on NSGA-II,'' \textit{Neurocomputing}, vol. 522, pp. 116--128, 2023.
	
	\bibitem{ding2024fmdnn} W. Ding, T. Zhou, J. Huang, S. Jiang, T. Hou, and C.-T. Lin, ``Fmdnn: A fuzzy-guided multigranular deep neural network for histopathological image classification,'' \textit{IEEE Transactions on Fuzzy Systems}, vol. 32, no. 8, pp. 4709--4723, 2024.

	\bibitem{lecun2002gradient}
	Y. LeCun, L. Bottou, Y. Bengio, and P. Haffner, ``Gradient-based learning applied to document recognition,'' \textit{Proceedings of the IEEE}, vol. 86, no. 11, pp. 2278--2324, 2002.
	
	\bibitem{he2016deep}
	K. He, X. Zhang, S. Ren, and J. Sun, ``Deep residual learning for image recognition,'' in \textit{Proceedings of the IEEE Conference on Computer Vision and Pattern Recognition}, 2016, pp. 770--778.

	\bibitem{vaswani2017attention} A. Vaswani, N. Shazeer, N. Parmar, J. Uszkoreit, L. Jones, A. N. Gomez, Ł. Kaiser, and I. Polosukhin, ``Attention is all you need,'' \textit{Advances in Neural Information Processing Systems}, vol. 30, 2017.
	
	\bibitem{battaglia2018relational} P. W. Battaglia et al., ``Relational inductive biases, deep learning, and graph networks,'' arXiv preprint arXiv:1806.01261, 2018.

	\bibitem{jiang2019semi} B. Jiang, Z. Zhang, D. Lin, J. Tang, and B. Luo, ``Semi-supervised learning with graph learning-convolutional networks,'' in \textit{Proc. IEEE/CVF Conf. Comput. Vis. Pattern Recognit.}, 2019, pp. 11313--11320.
		
	\bibitem{grover1996fast}
	L. K. Grover, ``A fast quantum mechanical algorithm for database search,'' in \textit{Proc. 28th Annu. ACM Symp. Theory Comput.}, 1996, pp. 212--219.
	
	\bibitem{shor1994algorithms}
	P. W. Shor, ``Algorithms for quantum computation: discrete logarithms and factoring,'' in \textit{Proc. 35th Annu. Symp. Found. Comput. Sci.}, 1994, pp. 124--134.
	
	\bibitem{wu2024quantum}
	S. Wu, R. Li, Y. Song, S. Qin, Q. Wen, and F. Gao, ``Quantum assisted hierarchical fuzzy neural network for image classification,'' \textit{IEEE Transactions on Fuzzy Systems}, 2024.
	
	\bibitem{tiwari2024quantum} P. Tiwari, L. Zhang, Z. Qu, and G. Muhammad, ``Quantum fuzzy neural network for multimodal sentiment and sarcasm detection,'' \textit{Information Fusion}, vol. 103, p. 102085, 2024.
	
	\bibitem{babu2025entanglement} A. Babu et al., ``Entanglement-enabled quantum kernels for enhanced feature representation,'' \textit{APL Quantum}, vol. 2, no. 1, 2025.
	
	\bibitem{preskill2018quantum}
	J. Preskill, ``Quantum computing in the NISQ era and beyond,'' \textit{Quantum}, vol. 2, p. 79, 2018.
	
	\bibitem{zadeh1973outline} L. A. Zadeh, ``Outline of a new approach to the analysis of complex systems and decision processes,'' \textit{IEEE Transactions on Systems, Man, and Cybernetics}, no. 1, pp. 28--44, 1973.
	
	\bibitem{lin1991neural} C.-T. Lin and C. S. G. Lee, ``Neural-network-based fuzzy logic control and decision system,'' \textit{IEEE Transactions on Computers}, vol. 40, no. 12, pp. 1320--1336, 1991.
	
	\bibitem{jang1993anfis} J.-S. R. Jang, ``ANFIS: adaptive-network-based fuzzy inference system,'' \textit{IEEE Transactions on Systems, Man, and Cybernetics}, vol. 23, no. 3, pp. 665--685, 1993.

	\bibitem{mohamed2025deep} N. Mohamed, R. L. Almutairi, S. Abdelrahim, R. Alharbi, F. M. Alhomayani, A. Alsulami, and S. Alkhalaf, ``Deep convolutional fuzzy neural networks with stork optimization on chronic cardiovascular disease monitoring for pervasive healthcare services,'' \textit{Scientific Reports}, vol. 15, no. 1, p. 19008, 2025.

	\bibitem{dosovitskiy2020image}
	A. Dosovitskiy, ``An image is worth 16x16 words: Transformers for image recognition at scale,'' \textit{arXiv preprint arXiv:2010.11929}, 2020.
	
	\bibitem{chen2025relgnn} T. Chen, C. Kanatsoulis, and J. Leskovec, ``Relgnn: Composite message passing for relational deep learning,'' arXiv preprint arXiv:2502.06784, 2025.

	\bibitem{wu2025large} F. Wu, V. P. Dwivedi, and J. Leskovec, ``Large language models are good relational learners,'' arXiv preprint arXiv:2506.05725, 2025.

	\bibitem{rebentrost2014quantum} P. Rebentrost, M. Mohseni, and S. Lloyd, ``Quantum support vector machine for big data classification,'' \textit{Phys. Rev. Lett.}, vol. 113, no. 13, p. 130503, 2014.

	\bibitem{qu2024quantum} Z. Qu, L. Zhang, and P. Tiwari, ``Quantum fuzzy federated learning for privacy protection in intelligent information processing,'' \textit{IEEE Transactions on Fuzzy Systems}, vol. 33, no. 1, pp. 278--289, 2024.	

	\bibitem{huang2022quantum} H.-Y. Huang et al., ``Quantum advantage in learning from experiments,'' \textit{Science}, vol. 376, no. 6598, pp. 1182--1186, 2022.
	
	\bibitem{yao2025hqfnn}
	J.~Yao and Y.~Guo, ``HQFNN: A compact quantum-fuzzy neural network for accurate image classification,'' \textit{arXiv preprint arXiv:2506.11146}, 2025.

	\bibitem{yerkin2025fuzzy}
	A.~Yerkin \textit{et al.}, ``Fuzzy theory in computer vision: A review,'' \textit{arXiv preprint arXiv:2507.18660}, 2025.

	\bibitem{zadeh1971similarity}
	L.~A. Zadeh, ``Similarity relations and fuzzy orderings,'' \textit{Information Sciences}, vol.~3, no.~2, pp. 177--200, 1971.
	
	\bibitem{zadeh1975concept}
	L.~A. Zadeh, ``The concept of a linguistic variable and its application to approximate reasoning—I,'' \textit{Information Sciences}, vol.~8, no.~3, pp. 199--249, 1975.
	
	\bibitem{atanassov1999intuitionistic}
	K.~T. Atanassov, ``Intuitionistic fuzzy sets,'' in \textit{Intuitionistic Fuzzy Sets: Theory and Applications}. Heidelberg, Germany: Springer, 1999, pp. 1--137.
	
	\bibitem{zimmermann1980latent}
	H.-J. Zimmermann and P. Zysno, ``Latent connectives in human decision making,'' \textit{Fuzzy Sets and Systems}, vol. 4, no. 1, pp. 37--51, 1980.
	
	\bibitem{schuld2021machine}
	M. Schuld and F. Petruccione, \textit{Machine Learning with Quantum Computers}. Springer, 2021, vol. 676.

	\bibitem{sim2019expressibility}
	S. Sim, P. D. Johnson, and A. Aspuru-Guzik, "Expressibility and entangling capability of parameterized quantum circuits for hybrid quantum-classical algorithms," \textit{Advanced Quantum Technologies}, vol. 2, no. 12, p. 1900070, 2019.

	\bibitem{lyons1998coding}
	M.~Lyons, S.~Akamatsu, M.~Kamachi, and J.~Gyoba, ``Coding facial expressions with {Gabor} wavelets,'' in \textit{Proc. 3rd IEEE Int. Conf. Autom. Face Gesture Recognit.}, 1998, pp. 200--205.
	
	\bibitem{lyons2021excavating}
	M.~J. Lyons, ``{`Excavating AI'} re-excavated: Debunking a fallacious account of the {JAFFE} dataset,'' \textit{arXiv preprint arXiv:2107.13998}, 2021.

	\bibitem{krizhevsky2009learning}
	A. Krizhevsky, G. Hinton, \textit{et al.}, ``Learning multiple layers of features from tiny images,'' Univ. Toronto, Toronto, ON, Canada, Tech. Rep., 2009.
	
	\bibitem{xiao2017fashion}
	H. Xiao, K. Rasul, and R. Vollgraf, ``Fashion-MNIST: A novel image dataset for benchmarking machine learning algorithms,'' \textit{arXiv preprint arXiv:1708.07747}, 2017.
	
	\bibitem{lazebnik2006beyond}
	S. Lazebnik, C. Schmid, and J. Ponce, ``Beyond bags of features: Spatial pyramid matching for recognizing natural scene categories,'' in \textit{Proc. IEEE Comput. Soc. Conf. Comput. Vis. Pattern Recognit. (CVPR)}, vol. 2, 2006, pp. 2169--2178.
	
	\bibitem{mitarai2018quantum}
	K. Mitarai, M. Negoro, M. Kitagawa, and K. Fujii, ``Quantum circuit learning,'' \textit{Phys. Rev. A}, vol. 98, no. 3, p. 032309, 2018.
	
	\bibitem{schuld2019evaluating}
	M. Schuld, V. Bergholm, C. Gogolin, J. Izaac, and N. Killoran, ``Evaluating analytic gradients on quantum hardware,'' \textit{Phys. Rev. A}, vol. 99, no. 3, p. 032331, 2019.

	\bibitem{loshchilov2017decoupled}
	I. Loshchilov and F. Hutter, ``Decoupled weight decay regularization,'' \textit{arXiv preprint arXiv:1711.05101}, 2017.
	
	\bibitem{wang2022torchquantum}
	H. Wang \textit{et al.}, ``Torchquantum case study for robust quantum circuits,'' in \textit{Proc. 41st IEEE/ACM Int. Conf. Comput.-Aided Design}, 2022, pp. 1--9.
	
	\bibitem{mukhoti2023deep}
	J. Mukhoti, A. Kirsch, J. Van Amersfoort, P. H. S. Torr, and Y. Gal, ``Deep deterministic uncertainty: A new simple baseline,'' in \textit{Proc. IEEE/CVF Conf. Comput. Vis. Pattern Recognit. (CVPR)}, 2023, pp. 24384--24394.
	
	\bibitem{kraskov2004estimating}
	A. Kraskov, H. St{\"o}gbauer, and P. Grassberger, ``Estimating mutual information,'' \textit{Phys. Rev. E}, vol. 69, no. 6, p. 066138, 2004.

	\bibitem{nielsen2010quantum}
	M. A. Nielsen and I. L. Chuang, \textit{Quantum Computation and Quantum Information}, 10th ed. Cambridge, U.K.: Cambridge University Press, 2010.


\end{thebibliography}

\end{document}